\documentclass{PoS}

\title{A rational approach to $\eta$ and $\eta^\prime$ transition form factors}

\ShortTitle{A rational approach to $\eta$ and $\eta^\prime$ transition form factors}

\author{
\speaker{Rafel Escribano}
\thanks{I would like to express my gratitude to the HADRON 13 Organizing Committee
for the opportunity of presenting this contribution, and for the pleasant and interesting workshop we have enjoyed.}\\
Grup de F\'{\i}sica Te\`orica (Departament de F\'{\i}sica) and Institut de F\'{\i}sica d'Altes Energies (IFAE),
Universitat Aut\`onoma de Barcelona, E-08193 Bellaterra (Barcelona), Spain\\
E-mail: \email{rescriba@ifae.es}
}

\abstract{
The $\eta$ and $\eta^\prime$ transition form factors in the space-like region are analyzed at low and intermediate energies
in a model-independent way through the use of rational approximants.
The slope and curvature parameters of the form factors as well as their values at infinity are extracted from experimental data.
The impact of these results on the mixing parameters of the $\eta$-$\eta^\prime$ system are also discussed.
}

\FullConference{
XV International Conference on Hadron Spectroscopy-Hadron 2013\\
4-8 November 2013\\
Nara, Japan }

\begin{document}

The $\eta$ and $\eta^\prime$ transition form factors (TFFs) encode the effects of the strong force 
in the interaction of two photons with the $\eta$ and $\eta^\prime$, respectively.
In the space-like region of the TFFs, the $\eta$ or $\eta^\prime$ are produced, for instance,
in the reaction $e^+e^-\to e^+e^-P$, with $P=\eta, \eta^\prime$.
Using this reaction, the energy dependence of the TFFs is measured in the following way:
one of the leptons is scattered at large angles emitting a highly virtual photon
and the other is little scattered thus producing a quasi-real photon.
The scattered lepton is then tagged while the other remains untagged.
This is the single-tag method which at present and before has been used to measured the TFFs,
not only for the $\eta$ and $\eta^\prime$ but also for the $\pi^0$.
Then, the energy dependence of the TFFs, which in principle is a function of the two photon virtualities
$F_{P\gamma^\ast\gamma^\ast}(q^2_1,q^2_2)$, reduces to $F_{P\gamma^\ast\gamma}(Q^2)$, with $Q^2=-q^2$,
for the case of a single-tagged lepton and in the space-like region of phase-space.

\begin{figure}[pt]
\centerline{
\includegraphics[width=0.5\textwidth]{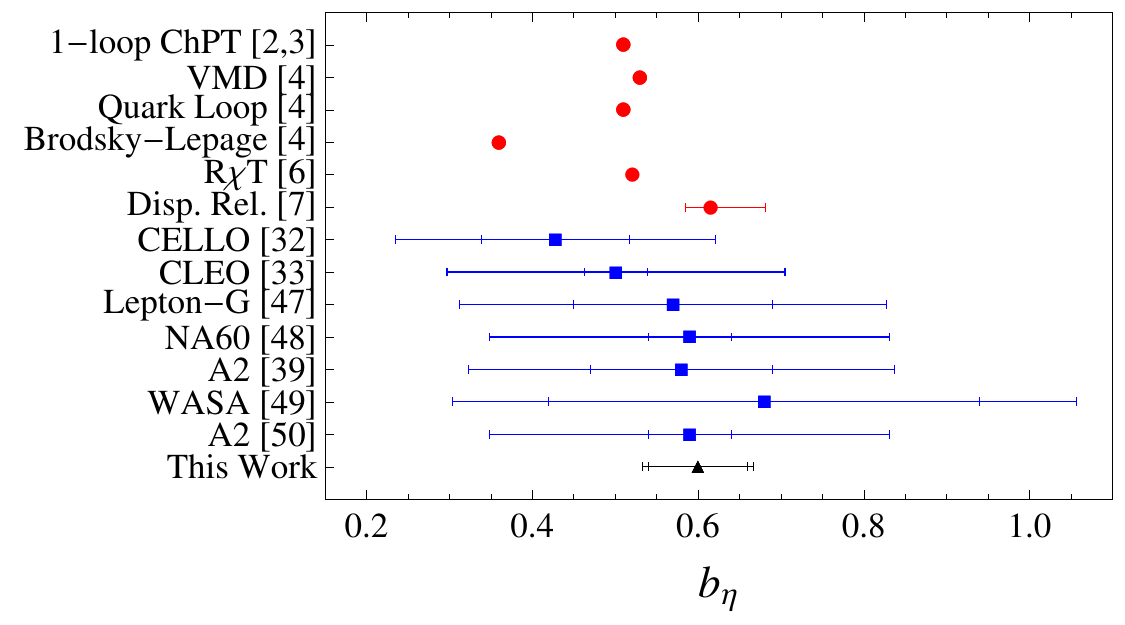}
\includegraphics[width=0.5\textwidth]{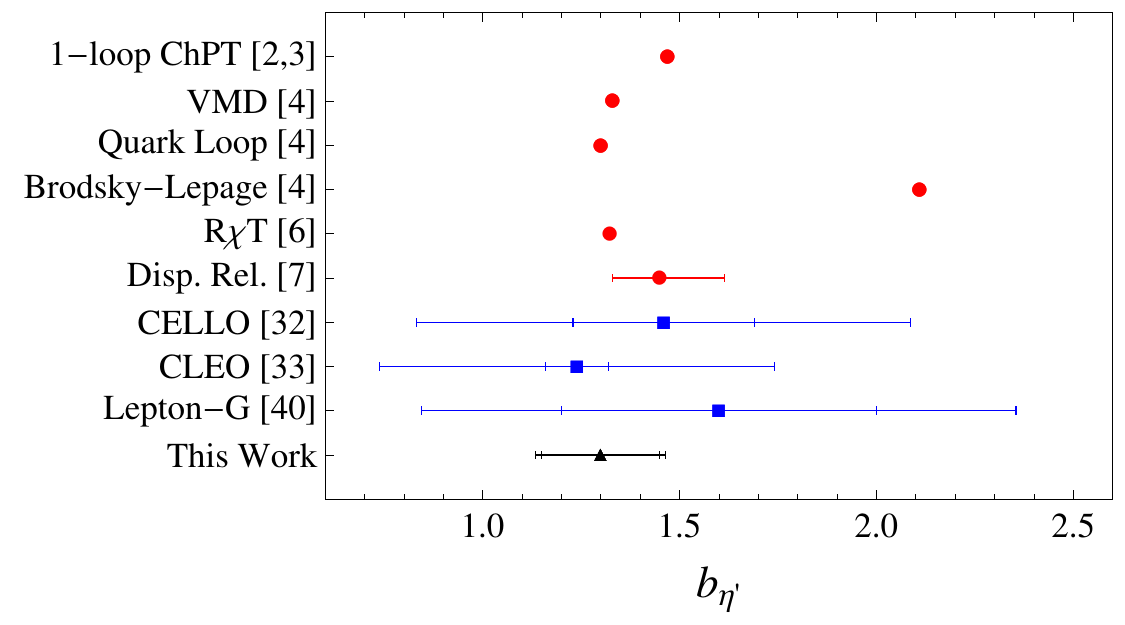}
}
\caption{Slope parameter for the $\eta$ (left) and $\eta^\prime$ (right) TFFs from different theoretical (red circles) and experimental (blue squares) 
determinations mentioned in the text.
Inner error is statistical and larger error is the combination of both statistical and systematic.
\label{fig:slopecomp}
}
\end{figure}

In the low-energy region, the TFF can be expanded as
\begin{equation}
\label{TFF0}
F_{P\gamma^\ast\gamma}(Q^2)=
F_{P\gamma\gamma}(0)\left(1-b_P\frac{Q^2}{m_P^2}+c_P\frac{Q^4}{m_P^4}+\cdots\right)\ ,
\end{equation}
where $F_{P\gamma\gamma}(0)$ is the normalization, 
the parameters $b_P$ and $c_P$ are the slope and curvature, respectively,
and $m_P$ is the pseudoscalar meson mass.
$F_{P\gamma\gamma}(0)$ can be fix from experiment through the measurement of
$\Gamma(\eta^{(\prime)}\to\gamma\gamma$),
\begin{equation}
\label{FPgg0}
|F_{P\gamma\gamma}(0)|^2=\frac{64\pi}{(4\pi\alpha)^2}\frac{\Gamma(P\to\gamma\gamma)}{m_P^3}\ ,
\end{equation}
or from theory by means of the prediction of the QCD axial anomaly in the chiral and large-$N_c$ limits simultaneously,
that is, $F_{\pi^0\gamma\gamma}(0)=1/(4\pi^2 F_\pi)$, with $F_\pi\simeq 92$ MeV, for the $\pi^0$, and
\begin{equation}
\label{FPgg0etaetap}
F_{\eta\gamma\gamma}(0)=\frac{1}{4\pi^2}\left(\frac{\hat c_q}{F_q}{\rm c}\phi-\frac{\hat c_s}{F_s}{\rm s}\phi\right)\ ,\quad
F_{\eta^\prime\gamma\gamma}(0)=\frac{1}{4\pi^2}\left(\frac{\hat c_q}{F_q}{\rm s}\phi+\frac{\hat c_s}{F_s}{\rm c}\phi\right)\ ,
\end{equation}
with $\hat c_q=5/3$, $\hat c_s=\sqrt{2}/3$ and $({\rm s,c})\equiv(\sin,\cos)$.
Concerning the slope parameter, there is a wide variety of approaches that predict it
(see Fig.~\ref{fig:slopecomp} for a comparison of numerical predictions):
Chiral Perturbation Theory (ChPT) \cite{Ametller:1991jv}, Vector Meson Dominance (VMD) \cite{Ametller:1991jv}, 
constituent-quark loops \cite{Ametller:1991jv}, the Brodsky-Lepage (BL) interpolation formula \cite{Brodsky:1981rp}, 
Resonance Chiral Theory (RChT) \cite{Czyz:2012nq}, and more recently, a dispersive analysis \cite{Hanhart:2013vba}.
On the experimental side, these parameters are usually obtained from a fit to data using
a normalized, single-pole term with an associated mass $\Lambda_P$, \textit{i.e.}
\begin{equation}
\label{TFFmonopole}
F_{P\gamma^\ast\gamma}(Q^2)=\frac{F_{P\gamma\gamma}(0)}{1+Q^2/\Lambda_P^2}\ .
\end{equation}
In this case, $b_P=m_P^2/\Lambda_P^2$ and $c_P=b_P^2$.

In the high-energy region, the TFF is expressed, within the framework of perturbative QCD (pQCD),
as a convolution of a perturbative hard scattering amplitude and the soft non-perturbative wave function of the meson \cite{Lepage:1980fj}.
The asymptotic behaviour of the TFFs in the limit $Q^2\to\infty$ is then given by
\begin{equation}
\label{TFFinfinity}
\begin{array}{l}
\displaystyle
\lim_{Q^2\to\infty} Q^2 F_{\eta\gamma^\ast\gamma}(Q^2)=
2(\hat c_q F_q{\rm c}\phi-\hat c_s F_s{\rm s}\phi)\ ,\\[2ex]
\displaystyle
\lim_{Q^2\to\infty} Q^2 F_{\eta^\prime\gamma^\ast\gamma}(Q^2)=
2(\hat c_q F_q{\rm s}\phi+\hat c_s F_s{\rm c}\phi)\ ,\
\end{array}
\end{equation}
in the same way as $\lim_{Q^2\to\infty} Q^2 F_{\pi^0\gamma^\ast\gamma}(Q^2)=2F_\pi$ for the $\pi^0$ (see Ref.~\cite{Brodsky:1981rp}).

While the low- and high-energy regions are in principle well described by ChPT and pQCD, respectively,
to have a model-independent description of the TFFs in the whole energy range is unfortunately still lacking for the $\eta$ and $\eta^\prime$.
In Ref.~\cite{Masjuan:2012wy}, it was suggested for the $\pi^0$ case that this model-independent approach can be achieved
using a sequence of rational functions, the Pad\'e Approximants (PAs), to fit the experimental data.
In this way, not only the low- and high-energy predictions of ChPT and pQCD should be reproduced but also a reliable description of the 
intermediate-energy region would be available.
The main advantage of the method of PAs is indeed to provide the $Q^2$ dependence of the TFF over the whole space-like region
in an easy, systematic and model-independent way \cite{Masjuan:2008fv}.
It is the purpose of the work in Ref.~\cite{Escribano:2013kba}, in which this proceeding is based on,
to present such a description for the $\eta$ and $\eta^\prime$ cases.

PAs are rational functions $P^N_M(Q^2)$
(ratio of a polynomial $T_N(Q^2)$ of order $N$ and a polynomial $R_M(Q^2)$ of order $M$)
constructed in such a way that they have the same Taylor expansion as the function to be approximated up to order 
${\cal O}(Q^2)^{N+M+1}$ (see Ref.~\cite{Pade} for details).
For certain special functions, there are convergence theorems that guarantee the convergence of different PA sequences to the given function \cite{Pade}.
For the case of the TFF, we don't know the function itself, nor its analytic structure in detail.
The only information we have, if we want to use it, is the asymptotic behaviour of that function in the low- and high-energy limits.
Moreover, due to the limited statistics of the experimental data, it will not be possible to build an infinite sequence of PAs.
The best we can do is to check whether several well-motivated sequences of PAs show an asymptotic behaviour when compared to 
the exact predictions of some given well-established models.
If so, we can be confident, even though it is not proven, that the proposed sequences of PAs will also converge to the unknown function behind the TFF.
It is in this sense that the PAs method has to be considered as a model-independent way of analyzing the TFFs in the whole energy region.
The issue of being the PAs a systematic procedure can be quantified in this case through a systematic error.
For some observable related to the TFF, such as the normalization or the slope and curvature parameters,
this error is defined as the relative error between the exact result for the observable in some given model and the prediction of some PA at some order.
Once this procedure is performed for several models and different sequences of PAs,
we choose as the systematic error associated to that particular observable
the worst of the relative errors that were found, thus being as much conservative as possible.
This systematic error will take into account the fact that we don't know the exact function of the TFF,
that we have used several different PAs sequences, and that we can reach only a finite order for these sequences.
However, the predictions obtained in this way are robust and do not depend on any physical model
(maybe only supplemented by some physical input if required for the lack of precision),
they have been just obtained from a mathematical treatment of experimental data.
Model-dependent predictions for the observables should lie, if the model is correct, inside the error windows allow by this procedure.

For the $\eta$ and $\eta^\prime$ TFFs, we use two different types of PA sequences, 
the $P^N_1(Q^2)$ (single-pole approximants) and $P^N_N(Q^2)$ (diagonal approximants).
The $P^N_1(Q^2)$ sequence seems the optimal choice
if an appropriate combination of the $\rho$, $\omega$ and $\phi$ mesons plays the same role, as an effective single-pole dominance,
as the $\rho$ on the $\pi^0$ TFF,
where this $\rho$ meson contribution exhibits a predominant role with the excited states being much suppressed.
According to Ref.~\cite{Lepage:1980fj}, the pseudoscalar TFFs behave as $1/Q^2$ for $Q^2\to\infty$.
Therefore, it would be desirable to incorporate this asymptotic limit information in the fits by considering also a $P^N_N(Q^2)$ sequence.
As stated, the length of the sequences will be fixed by the limited statistics of the experimental data.
The systematic error for the slope and curvature parameters obtained from these two types of sequences is discussed in detail in
Ref.~\cite{Escribano:2013kba}.
The experimental data for the $\eta$ and $\eta^\prime$ TFFs in the space-like region come from the 
CELLO \cite{Behrend:1990sr}, CLEO \cite{Gronberg:1997fj}, \textit{BABAR} \cite{BABAR:2011ad}, 
and L3 Collabs.~\cite{Acciarri:1997yx}, the latter only for the $\eta^\prime$ case.
The $Q^2$ set of data ranges from the 0.62 GeV$^2$ of CELLO to the 34.38 GeV$^2$ of \textit{BABAR}, for the $\eta$,
and from the 0.0 GeV$^2$ of L3 to the 34.32 GeV$^2$ of \textit{BABAR}, for the $\eta^\prime$, respectively.
Since it is common to present data in the form of $Q^2|F_{P\gamma^\ast\gamma}(Q^2)|$ instead of $|F_{P\gamma^\ast\gamma}(Q^2)|$,
we prefer to fit the first form.
In Ref.~\cite{Escribano:2013kba}, many different fits are discussed.
However, in this proceeding, we only present the best ones, fits were two supplementary conditions are imposed.
First, $\lim_{Q^2\to 0} Q^2 F_{\eta^{(\prime)}\gamma^\ast\gamma}(Q^2)=0$, 
a condition which is satisfied because the TFFs are known to be non-singular at the origin.
Second, the normalisations $F_{\eta^{(\prime)}\gamma\gamma}(0)$
will be fixed, through Eq.~(\ref{FPgg0}), to the experimental values obtained from the measurements of the respective two-photon partial widths.
We use $\Gamma_{\eta\to\gamma\gamma}=0.516(18)$ keV,
after combining the PDG average \cite{Beringer:1900zz} together with the recent KLOE-2 result \cite{Babusci:2012ik}, and 
$\Gamma_{\eta^\prime\to\gamma\gamma}=4.35(14)$ keV from the PDG fit \cite{Beringer:1900zz}.

\begin{table}[pt]
\centering
{\begin{tabular}{ccccccccc}
\hline
& $N$ & $b_{\eta}$ & $c_{\eta}$ & $\chi^2$/dof & $N$ & $b_{\eta^\prime}$ & $c_{\eta^\prime}$ & $\chi^2$/dof\\
\hline
$P^N_1$ & $5$ & $0.58(6)$ & $0.34(8)$ & $0.80$ & $6$ & $1.30(15)$ & $1.72(47)$ & $0.70$\\[0.5ex]
$P^N_N$ & $2$ & $0.66(10)$ & $0.47(15)$ & $0.77$ & $1$ & $1.23(3)$ & $1.52(7)$ & $0.67$\\[0.5ex]
Final & $ $ & $0.60(6)$ & $0.37(10)$ & $ $ & $ $ & $1.30(15)$ & $1.72(47)$ & $ $\\
\hline
\end{tabular}
\caption{$\eta$ (left) and $\eta^\prime$ (right) slope and curvature parameters
obtained from the best fits to experimental data \emph{including} the measured two-photon partial decay widths.
The first column indicates the type of sequence used for the fit and $N$ is the highest order reached for that sequence.
The last row shows the weighted average result for each low-energy parameter.}
\label{tab:psresults2}
}
\end{table}

Our results for the $\eta$ and $\eta^\prime$ slope and curvature parameters obtained from these best fits are shown in Table \ref{tab:psresults2}.
The quality of the fits are similar in all cases.
The values of the low-energy parameters (LEPs) for each meson are in agreement within errors
(only symmetrized statistical errors are shown).
The difference between the single-pole approximants, $P^N_1$, and the diagonal ones, $P^N_N$,
is that the former do not include the asymptotic behaviour expected from pQCD which is imposed in the latter.
For the $\eta^\prime$ case, the weighted average is not performed because the systematic errors of the LEPs
obtained from the $P^1_1$ approximant are large and thus considered as unacceptable
(see Ref.~\cite{Escribano:2013kba} for details).
Our final results are
\begin{equation}
\label{finalresults}
\begin{array}{l}
b_\eta=0.60(6)_{\rm stat}(3)_{\rm sys}\ ,\\[1ex]
c_\eta=0.37(10)_{\rm stat}(7)_{\rm sys}\ ,
\end{array}
\qquad
\begin{array}{l}
b_{\eta^\prime}=1.30(15)_{\rm stat}(7)_{\rm sys}\ ,\\[1ex]
c_{\eta^\prime}=1.72(47)_{\rm stat}(34)_{\rm sys}\ ,
\end{array}
\end{equation}
where the second uncertainty is the most conservative systematic error in each case
(of the order of 5\% and 20\% for $b_P$ and $c_P$, respectively).
A comparison of the former results for the slope parameter of both mesons
and different theoretical and experimental determinations found in the literature
are shown in Fig.~\ref{fig:slopecomp}.

The $\eta$-$\eta^\prime$ mixing parameters in the quark-flavour basis are obtained from
Eqs.~(\ref{FPgg0etaetap}) and (\ref{TFFinfinity}).
As an input, we use the normalization at zero of both TFFs, 
$|F_{\eta\gamma\gamma}(0)|=0.274(5)$ GeV$^{-1}$ and $|F_{\eta^\prime\gamma\gamma}(0)|=0.344(6)$ GeV$^{-1}$,
from the measured decay widths,
and for the asymptotic value of the $\eta$ TFF we take the predicted value
$\lim_{Q^2\to\infty}Q^2 F_{\eta\gamma^\ast\gamma}(Q^2)=0.160(24)$ GeV.
With these values, the mixing parameters are predicted to be
\begin{equation}
\label{mixingpar}
F_q/F_\pi=1.06(1)\ ,\qquad
F_s/F_\pi=1.56(24)\ ,\qquad
\phi=40.3(1.8)^\circ\ ,
\end{equation}
with $F_\pi=92.21(14)$ MeV \cite{Beringer:1900zz}.
They can be compared, for instance, with the mixing parameters obtained in Ref.~\cite{Escribano:2005qq},
$F_q/F_\pi=1.10(3)$, $F_s/F_\pi=1.66(6)$ and $\phi=40.6(0.9)^\circ$,
after a careful analysis of $V\to\eta^{(\prime)}\gamma$, $\eta^{(\prime)}\to V\gamma$, with $V=\rho, \omega, \phi$,
and  $\eta^{(\prime)}\to\gamma\gamma$ decays, and the ratio
$R_{J/\psi}\equiv\Gamma(J/\psi\to\eta^\prime\gamma)/\Gamma(J/\psi\to\eta\gamma)$.
The agreement between this determination and the values in Eq.~(\ref{mixingpar}) is quite impressive since
only the information of the TFFs is used to predict these mixing parameters.

\end{document}